\documentclass[conference]{IEEEtran}

\usepackage{comment}
\IEEEoverridecommandlockouts
\usepackage{cite}
\usepackage{amsmath,amssymb,amsfonts}
\usepackage{algorithmic}
\usepackage[linesnumbered,ruled,vlined]{algorithm2e}
\usepackage{graphicx}
\usepackage{textcomp}
\usepackage{xcolor}
\usepackage[T1]{fontenc}
\usepackage{subfigure}
\usepackage{multirow}
\usepackage{booktabs}
\usepackage{enumitem}
\usepackage{pifont}
\usepackage{graphicx}
\usepackage{wrapfig}
\usepackage[labelfont=bf,font=small]{caption}
\def\BibTeX{{\rm B\kern-.05em{\sc i\kern-.025em b}\kern-.08em
    T\kern-.1667em\lower.7ex\hbox{E}\kern-.125emX}}

\usepackage[bookmarks=true,breaklinks=true,colorlinks,linkcolor=blue,citecolor=blue,urlcolor=blue]{hyperref}

\usepackage{comment}
\usepackage{xspace}
\newcommand{\design}{{CopyQNN}\xspace}

\begin{document}
\title{\design: Quantum Neural Network Extraction Attack under Varying Quantum Noise}

\author{
\centering
\begin{tabular}{ccccccc}
Zhenxiao Fu$^*$   &
Leyi Zhao$^*$     &
Xuhong Zhang$^*$  & 
Yilun Xu$^\dag$   &
Gang Huang$^\dag$ &
Fan Chen$^*$ \\
\multicolumn{3}{c}{$^*$Indiana University Bloomington} & 
\multicolumn{3}{c}{$^\dag$Lawrence Berkeley National Laboratory}\\
\multicolumn{3}{c}{$^*$\{zhfu, leyizhao, zhangxuh, fc7\}@iu.edu} & 
\multicolumn{3}{c}{$^\dag$\{yilunxu, ghuang\}@lbl.gov}\\
\end{tabular}
\vspace{-24pt}
}

\maketitle

\begin{abstract}
Quantum Neural Networks (QNNs) have shown significant value across domains, with well-trained QNNs representing critical intellectual property often deployed via cloud-based QNN-as-a-Service (QNNaaS) platforms. Recent work has examined QNN model extraction attacks using classical and emerging quantum strategies. These attacks involve adversaries querying QNNaaS platforms to obtain labeled data for training local substitute QNNs that replicate the functionality of cloud-based models.
However, existing approaches have largely overlooked the impact of varying quantum noise inherent in noisy intermediate-scale quantum (NISQ) computers, limiting their effectiveness in real-world settings.
To address this limitation, we propose the~\design~framework, which employs a three-step data cleaning method to eliminate noisy data based on its noise sensitivity. This is followed by the integration of contrastive and transfer learning within the quantum domain, enabling efficient training of substitute QNNs using a limited but cleaned set of queried data.
Experimental results on NISQ computers demonstrate that a practical implementation of~\design~significantly outperforms state-of-the-art QNN extraction attacks, achieving an average performance improvement of 8.73\% across all tasks while reducing the number of required queries by 90$\times$, with only a modest increase in hardware overhead.
\end{abstract}


\IEEEpeerreviewmaketitle


\section{Introduction}
Quantum Neural Networks (QNNs) are promising noisy intermediate-scale quantum (NISQ) algorithms, known for their ability to solve complex problems on current noisy hardware. 
However, their development requires domain expertise~\cite{schuld2015introduction} and significant optimization~\cite{Zhu:SCIENCE2019}, making QNNs valuable intellectual properties (IPs) that demand stringent protection.
QNNs are currently deployed as QNN-as-a-Service (QNNaaS)~\cite{Google, QAI} in the cloud, as illustrated in Figure~\ref{f:background_qnn}, where users interact with QNNaaS by submitting queries with their local data. 
The server then executes the QNN on its NISQ devices, measures the outputs, and provides classical probability results to end users for subsequent post-processing, ultimately yielding a classical prediction.

Due to the high costs and limited availability of NISQ computing resources, the QNNaaS development model has become an attractive target for adversaries. These adversaries employ model extraction attacks~\cite{tramer2016stealing, Jacson:IJCNN2018, yu2020cloudleak, Fu:IJCNN2024, chen2025coteaching} to replicate a victim QNN by repeatedly querying the cloud model and using the obtained data to train a substitute. Once developed, the adversaries can exploit the stolen IP by querying the local substitute an unlimited number of times at no cost.

However, applying classical model extraction schemes~\cite{tramer2016stealing, Jacson:IJCNN2018, yu2020cloudleak} to QNNs yields low accuracy due to their failure to account for quantum noise~\cite{Huo:NJP2017, Clerk:RMP2010, Lax:PR1966}, a defining characteristic of NISQ computers. 
The recent QuantumLeak~\cite{Fu:IJCNN2024} (abbreviated as \textit{QLeak} in figures and tables for brevity) seeks to address this limitation by training an ensemble of substitute QNNs and leveraging their collective votes for final predictions to mitigate the impact of quantum noise. However, our preliminary results reveal that QuantumLeak struggles to account for the inherent fluctuations of practical NISQ computers. Furthermore, it requires a substantial number of queried data points (e.g., 6,000), leading to significantly increased costs (e.g., \$96 per minute for access to IBM quantum computers~\cite{QIBM}). This excessive querying not only raises the financial burden but also compromises stealth, as it risks triggering suspicion and alerting the QNNaaS provider, thereby jeopardizing the attack.
In this paper, we highlight these challenges and present our solutions and contributions, summarized as follows:

\begin{figure}[t!]
\centering
\includegraphics[width=\linewidth]{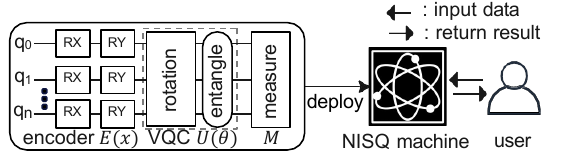}
\vspace{-0.2in}
\caption{The QNN-as-a-service on a NISQ server.}
\vspace{-0.2in}
\label{f:background_qnn}
\end{figure}

\begin{itemize}[leftmargin=*, nosep, topsep=2pt, partopsep=2pt]

\item \textbf{Varying quantum noise renders prior methods ineffective}.
Our preliminary results, which increased the number of query rounds to better capture the varying quantum noise in the NISQ devices, revealed a significant reduction in accuracy in the substitute QNN obtained by QuantumLeak. This suggests that the ensemble learning approach~\cite{ferreira2012boosting} used by QuantumLeak, which incorporates all noisy supervisory data, is less effective under these conditions. These findings highlight the need for a more robust method to manage noisy labels, one that rigorously filters out severely misleading data. By excluding these unreliable labels, the effectiveness of the QNN extraction process can potentially improve.

\item \textbf{Data cleaning is essential but reduces available training data}.
Filtering noisy supervisory data is crucial for training an accurate substitute QNN, but it reduces available training data and may impact performance. Moreover, maintaining the stealth of the attack imposes strict limitations on the number of queries, further constraining the data available for training.  
To address these challenges, we propose the~\design~framework. This approach uniquely combines contrastive learning\cite{simclr, moco, barlow}, which trains a quantum model to extract meaningful representations from source datasets, with transfer learning~\cite{oquab2014learning, radford2021learning}, which leverages the pre-trained feature extractor and trains only a small quantum classifier using the limited cleaned queried data. This integration within the quantum domain represents a novel approach that has not been previously explored.

\item \textbf{Limited NISQ resources and the need to preserve attack stealth}. 
We present NISQ implementation and training optimization for the~\design~framework. Through a detailed analysis of design costs and experimental results on NISQ computers, we demonstrate that a practical implementation of the~\design~architecture outperforms the state-of-the-art QNN extraction attack, QuantumLeak.
Specifically, \design~achieved an average improvement of 8.73\% across all evaluated tasks, while dramatically reducing the number of required queries by 90$\times$. 
These results underscore the effectiveness of the~\design~framework in leveraging limited quantum resources for enhanced QNN extraction attack performance.
\end{itemize}

\section{Background and Related Work}

\subsection{Key Concepts in Quantum Neural Networks}

\textbf{Basics of QNNs}.
QNNs are NISQ algorithms~\cite{bharti2022noisy} designed for noisy devices with inherent error tolerance.
As shown in Figure~\ref{f:background_qnn}, a QNN is implemented as quantum gates (e.g., \texttt{RX}, \texttt{RY}) executed on properly initialized qubits (e.g., $q_0$$\sim$$q_n$). For algorithms processing classical data, an encoder $E(\mathbf{x})$ embeds the input $\mathbf{x}$ into a quantum state, processed by a multilayer Variational Quantum Circuit (VQC), $U(\mathbf{\theta})$. The resulting quantum states are converted to raw probability vectors via a measurement layer $\mathbf{M}$, and a softmax function generates the final predicted label. Like classical models, QNN parameters (e.g., $\theta$) are trained using an optimizer~\cite{Zhu:SCIENCE2019}. Once developed, QNNs can be deployed on quantum cloud servers through QNN-as-a-Service (QNNaaS)~\cite{pennylane, Google, QAI}.
\textit{Creating a QNN requires domain expertise and costly data acquisition, making QNNs valuable intellectual property (IP) that necessitates robust protection measures.}

\textbf{Quantum Noises}.
NISQ devices are prone to various noise sources, including control imperfections~\cite{Huo:NJP2017}, crosstalk interference~\cite{Clerk:RMP2010}, and leakage~\cite{Lax:PR1966}.
Some of these noise sources are time-dependent, with their intensity varying over time. For example, in superconducting NISQ systems, fluctuations in unpaired electron populations can cause significant temporal changes in the decoherence rate~\cite{Gustavsson:SCI2016}, leading to variations in device parameters.
Table~\ref{tab:background_error} highlights parameter variations observed on the \texttt{IBM\_Brisbane} NISQ computer. Key metrics, including \textit{$T_1$} and \textit{$T_2$} coherence times, readout error rates, error rates for one-qubit gates (e.g., 1Q-Gate) and two-qubit gates (e.g., 2Q-Gate), and state preparation and measurement (SPAM) errors, were recorded at 6:00 and 18:00 on June 30, 2024. These measurements reveal notable discrepancies, with parameter fluctuations that may potentially result in a significant number of unpredictable erroneous labels during QNN queries conducted via QNNaaS.
\textit{Despite these impacts, particularly in the context of QNN model extraction attacks, prior research has not adequately explored this issue.}


\vspace{-0.12in}
\subsection{Related Work}
\vspace{-0.04in}

\textbf{Model Extraction Attacks}.
This work investigates model extraction attacks that query cloud services to label unlabeled data, which is then used to train a local substitute model, as established in prior research~\cite{tramer2016stealing, Jacson:IJCNN2018, yu2020cloudleak, Fu:IJCNN2024, chen2025coteaching}.  
Classical approaches~\cite{tramer2016stealing, Jacson:IJCNN2018, yu2020cloudleak} focus on traditional neural network extraction and overlook quantum noise in QNNs.
QuantumLeak~\cite{Fu:IJCNN2024} specifically addresses QNN extraction on NISQ devices using three main phases: (1) victim QNN query and data bagging, (2) QNN ensemble initialization and training, and (3) decision fusion through majority voting. However, our experiments reveal significant challenges for QuantumLeak in real-world environments.  
First, when we increased the number of query rounds from 3 (as in~\cite{Fu:IJCNN2024}) to 5 to account for varying quantum noise, the accuracy of the trained substitute QNN dropped substantially. Second, QuantumLeak’s requirement of 6000 queried data points imposes high costs, given the significant expense of NISQ computer access (e.g., \$96 per minute for IBM quantum computers~\cite{QIBM}). Moreover, excessive querying could raise suspicion, potentially alerting the QNNaaS provider and compromising the stealth of the attack.

\begin{table}[t!]
\centering
\captionof{table}{Error rates of \texttt{IBM\_Brisbane} on June/30/2024.}
\vspace{-0.06in}
\setlength{\tabcolsep}{3pt}
\begin{tabular}{|c|c|c|}
\hline
\textbf{Parameter}  & \textbf{06:00 (qubit 2/3)} & \textbf{18:00 (qubit 2/3)} \\\hline\hline
T1, T2 ($\mu$s)     & 223.5/137.5, 220.1/139.8      & 219.8/140.6, 223.8/138.5\\\hline 
Readout Error Rate       & 0.0123/0.0144         & 0.0142/0.0091\\\hline
1Q-Gate Error Rate       & 1.973e-4/2.144e-4     & 1.786e-4/2.36e-4 \\\hline
2Q-Gate Error Rate       & 4.56e-3               & 4.98e-3 \\\hline
Prob\_Meas0\_Prep1  & 0.0124/0.0082         & 0.0168/0.0082 \\\hline 
Prob\_Meas1\_Prep0  & 0.0074/0.0078         & 0.0116/0.0100 \\\hline
\end{tabular}
\label{tab:background_error}
\vspace{-0.2in}
\end{table}

\textbf{Learning from Noisy Labels}.
Existing methods for training neural network models with noisy labels typically fall into two main categories:  
(1) Including noisy supervision data while mitigating its impact:
This approach re-weights weaker instances using techniques such as ensemble learning~\cite{ferreira2012boosting}, selective training~\cite{ren2018learning}, or meta-learning~\cite{shu2019meta}. QuantumLeak has demonstrated the use of bagging-based ensemble construction~\cite{ferreira2012boosting} for QNN extraction attacks.  
(2) Excluding noisy supervision data:  
This involves employing data-cleaning algorithms to filter out noisy data~\cite{han2018coteaching, jiang2018mentornet}. Recent work~\cite{chen2025coteaching} has adapted the classical co-teaching method~\cite{han2018coteaching} to enhance QNN extraction attacks.


\textbf{Learning with Limited Data}.
Learning from limited data often leads to overfitting and poor generalization. To mitigate this issue, research has focused on two key approaches:  
(1) Transfer learning, which utilizes feature extractors pre-trained on large datasets from a source domain to improve performance on a target domain~\cite{oquab2014learning, radford2021learning}.  
(2) Contrastive learning, which enables models to capture meaningful representations from small target domain datasets by distinguishing augmented versions of the same instance from other instances~\cite{simclr, moco, barlow}.  
Recent studies have demonstrated the effectiveness of quantum transfer learning~\cite{wang2023hybrid} and quantum contrastive learning~\cite{jaderberg2022quantum} in classification tasks.  
Building on these successes, this work integrates transfer learning and contrastive learning within a QNN architecture to develop robust methods for QNN extraction attacks.


\begin{figure}[t!]
\centering
\includegraphics[width=1\linewidth]{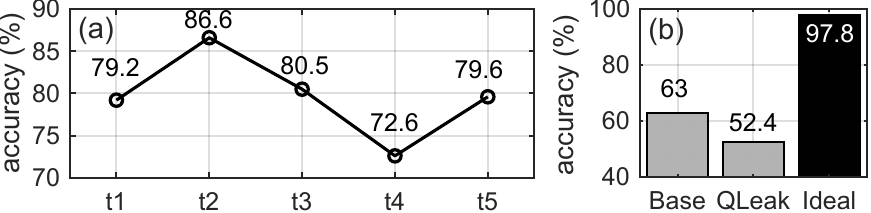}
\vspace{-0.18in}
\caption{Preliminary results indicate:  
(a) fluctuations in inference accuracy when running the victim QNN at different times during the day;
(b) ineffectiveness of existing model extraction attacks.}
\label{f:motivation-data-ab}
\vspace{-0.15in}
\end{figure}

\section{Preliminary Study}
\vspace{-0.03in}

We conduct a preliminary study to identify the overlooked impact of varying noise on QNN performance and the ineffectiveness of existing QNN extraction attack techniques. 
Detailed experimental configurations are provided in Section~\ref{sec:setup}.

\textbf{Accuracy Fluctuation on QNNaaS}. 
We train a victim QNN model for binary classification on the MNIST dataset using the $L2$ QNN architecture described in QuantumLeak. The victim model consists of a 4-qubit circuit with an amplitude encoding layer, two repeated VQC layers, and a measurement layer.
Simulation results without NISQ noise achieve an ideal binary classification accuracy of 97.8\%. However, when evaluated on the \texttt{IBM\_Brisbane} NISQ computer at five evenly spaced intervals throughout the day, the performance exhibits significant accuracy fluctuations, as shown in Figure~\ref{f:motivation-data-ab}(a). These fluctuations highlight the impact of quantum noise in QNNaaS queries conducted via the NISQ cloud, potentially resulting in noisy and unreliable labels.  
While QNNaaS platforms generally offer more stable accuracy due to cloud-side post-processing and error mitigation, the lack of detailed hardware information about the servers prevents users from fully understanding the factors contributing to this stability. 
\textit{This limitation complicates the training of an accurate substitute model and underscores the inherent challenges of leveraging data generated by NISQ devices.}

\textbf{Ineffectiveness of Existing Techniques}.
To accommodate practical NISQ conditions, we increased the number of query rounds from 3, as used in QuantumLeak, to 5, distributing them evenly across five intervals throughout the day.  All other settings, such as the total number of queried data points, ensemble committee size, and training configurations, were kept identical to those in QuantumLeak. We implemented the QuantumLeak technique, referred to as \textit{QLeak}, for QNN extraction.  We also developed a \textit{base} model that utilizes all queried data to train a local substitute QNN with an architecture identical to the victim QNN.
As shown in Figure~\ref{f:motivation-data-ab}(b), the inference accuracy of the local substitute models was compared against the ideal accuracy of the victim cloud QNN.  
Surprisingly, \textit{QLeak} exhibited reduced accuracy compared to the naive \textit{base} model, rather than achieving improvement.  
This observation indicates that noise fluctuations significantly degrade the quality of the queried labels, rendering the ensemble learning-based QuantumLeak scheme ineffective.  
\textit{The noisy supervision misleads the substitute model, emphasizing the need to first filter out excessively noisy labels and subsequently develop more effective techniques to train an accurate substitute model with reduced training data.}

\begin{figure}[t!]
\centering
\includegraphics[width=\linewidth]{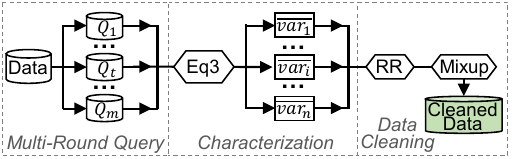}
\vspace{-0.15in}
\caption{Multi-round query, noise characterization, and data cleaning.}
\vspace{-0.1in}
\label{f:data_clearning}
\vspace{-0.05in}
\end{figure}

\vspace{-0.05in}
\section{\design}
\vspace{-0.05in}

To address the limitations of prior methods in accounting for practical variations in quantum noise, we propose the~\design~framework. The framework begins by filtering noise-prone queried data through a data-cleaning pipeline. It then combines contrastive learning with transfer learning to efficiently train a local substitute QNN using the refined, albeit smaller, training dataset.  
We first outline the threat model, followed by a detailed explanation of the core techniques. Finally, we present an NISQ implementation of the proposed framework and provide a comparative analysis of its implementation costs against QuantumLeak.

\vspace{-0.04in}
\subsection{Threat Model}
In the QNNaaS paradigm, users submit input data to a NISQ cloud machine, which returns raw probability output vectors. The client then applies classical operations (e.g., \texttt{softmax}) to produce the final predicted labels. For economic and privacy reasons, the details of the QNN architecture are kept confidential.  

\textbf{Attacker's Knowledge}.  
We adopt a black-box threat model, widely used in prior model extraction attacks~\cite{tramer2016stealing, Jacson:IJCNN2018, yu2020cloudleak, Fu:IJCNN2024}. In this model, the attacker has no access to any details of the victim QNN, including its training data, architecture, gate parameters, hyperparameters, or loss function. However, the attacker is assumed to have general domain knowledge of the training data, such as whether it pertains to images or text. The attacker can query the QNNaaS to obtain raw probability vectors, but excessive or anomalous querying may raise suspicion and alert the QNNaaS, compromising the stealth of the attack.  

\textbf{Attacker's Goals}.  
The attacker seeks to collect input-output pairs through queries to train a substitute QNN. This substitute model is intended to replicate the performance of the cloud QNN, enabling the attacker to perform unlimited queries locally without incurring additional costs.

\vspace{-0.04in}
\subsection{Query, Noise Characterization, and Data Cleaning}
Figure~\ref{f:data_clearning} outlines the three-step approach to address varying quantum noise. 
First, we conduct \textit{multi-round queries} to account for noise fluctuations. 
Next, we perform \textit{noise characterization} to identify susceptible data. 
Finally, we apply \textit{data cleaning} to filter out noise-prone data. Each of these steps is detailed in the following discussion.

\textbf{Multi-Round Query}.
\design~adopts the methodology from prior research~\cite{tramer2016stealing, Jacson:IJCNN2018, yu2020cloudleak, Fu:IJCNN2024} by constructing an unlabeled dataset from public sources and querying victim QNNs via QNNaaS.
To ensure comprehensive evaluation across varying noise levels on the NISQ server, \design~distributes queries evenly over 24 hours instead of clustering them within a short time frame.
Specifically, \design~schedules the data submission to the victim QNN in $m$ rounds (e.g., $m$=5) at intervals of $24/m$ hours.

\textbf{Noise Characterization}.
For the $n$ data samples sent to the QNNaaS platform for $d$-class inferences at round $t$ ({$1$}$\leq${$t$}$\leq${$m$}), the raw probability output for the $i_{th}$ data sample $I_i$ ({$1$}$\leq${$i$}$\leq${$n$}) can be represented as a $d$-dimensional vector:
\vspace{-6pt}\begin{equation}\vspace{-4pt}
Q_t(I_i)=[p_{t,i,1},\ldots, p_{t,i,j},\ldots, p_{t,i,d}]
\label{e:quantum_prob_all}
\end{equation}
where $p_{t,i,j}$ indicates the obtained probability of $I_i$ belonging to class $j$
({$1$}$\leq${$j$}$\leq${$d$}) at round $t$. 
The predicted label for data $I_i$ is typically computed as:
\vspace{-6pt}\begin{equation}\vspace{-4pt}
label_{t,i}=MI(SF(Q_t(I_i)))
\label{e:quantum_prob_label}
\end{equation}
where \texttt{SF}($\cdot$) represents the \texttt{softmax} function applied to \texttt{$Q_t$}($I_i$), and \texttt{MI} returns the index of the maximum element in the $d$-dimensional vector, which corresponds to the label.

In this work, we characterize the noisy inference output and then develop data cleaning schemes. Specifically, we construct an $m$-dimensional vector $P_{i,j}$=$[p_{1,i,j}, \ldots, p_{m,i,j}]$ for each data sample $I_i$. This vector contains the $m$ probabilities of the sample belonging to class $j$ from different query rounds. We then calculate the variance of these $m$ rounds of inference as:
\vspace{-8pt}\begin{equation}\vspace{-4pt}
\overline{var_i}=\frac{\sum_{j=1}^d var(P_{i,j})}{d}
\label{e:quantum_prob_variance}
\end{equation}
where \texttt{var} is a function returning the variance of a vector.

\begin{figure}[t!]
\centering
\includegraphics[width=\linewidth]{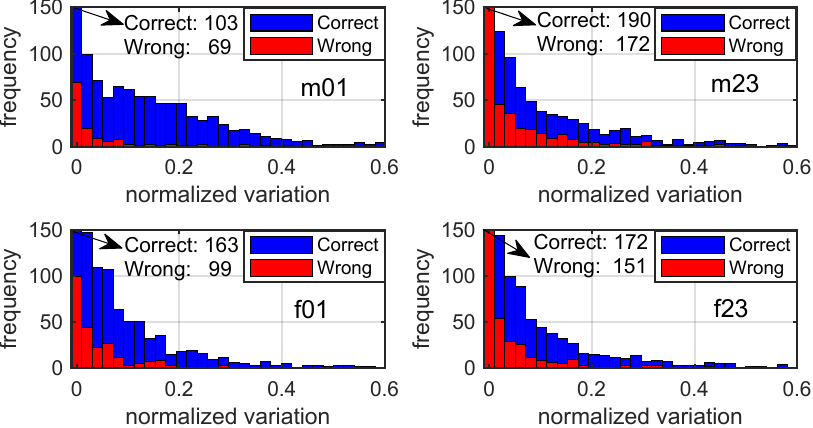}
\vspace{-0.2in}
\caption{Variance of queried labels over 5 rounds across different tasks.}
\vspace{-0.2in}
\label{f:var}
\end{figure}

\textbf{Data Cleaning}.
As illustrated in Figure~\ref{f:var}, we calculate the variance for $m$=5 query rounds, comparing the predicted labels with the ground truth. 
The results consistently demonstrate that mislabeled samples tend to cluster in regions with low variance, whereas correctly labeled samples are more broadly distributed. This pattern suggests that low variance is indicative of consistently erroneous labels, likely due to quantum noise. These findings highlight the potential of variance as a robust metric for identifying and filtering noise-prone samples to enhance data reliability.
To operationalize this insight, we propose a Remember Ratio (RR) to systematically filter low-variance samples. For example, with \texttt{RR}=0.2, the labeled data are ranked by variance in descending order, retaining only the top 20\%. This method effectively removes low-variance samples, which are more prone to mislabeling, thereby improving overall data quality by reducing noise.
Additionally, we apply Mixup~\cite{zhang2017mixup} to enhance the quality of the remaining queried labels. Overall, results demonstrate that \texttt{RR} significantly enhances the effectiveness of the design.
For instance, in the four tasks illustrated in Figure~\ref{f:var}, setting \texttt{RR}=0.6 can increase the proportion of clean samples from 
87.1\%, 63.1\%, 76\%, 68.1\% to 
96\%, 69.7\%, 83.5\%, 77.2\%, 
respectively. Comprehensive results with various \texttt{RR} values are provided in Section~\ref{sec:results}.



\begin{figure*}[t!]
\begin{minipage}{0.7\linewidth}
\begin{center}
\includegraphics[width=1\linewidth]{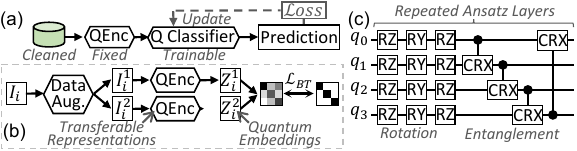}
\vspace{-0.2in}
\captionof{figure}{\design~overview: 
(a) the overall architecture and training of \texttt{QClassifier};
(b) the training process for transferable~\texttt{QEnc};
(c) the NISQ implementation of the VQC circuit.}
\label{f:overall_arch}
\end{center}
\end{minipage}
\hspace{0.02in}
\begin{minipage}{0.32\linewidth}
\setlength{\tabcolsep}{6pt}
\centering
\footnotesize
\begin{tabular}{|l|c|c|}
\hline
& \textbf{QLeak~\cite{Fu:IJCNN2024}} & \textbf{\design} \\\hline\hline
Data Size           & 6,000     &   40    \\\hline 
Query Rounds        & 3         &   5    \\\hline 
Query Lable \#      & 18, 000   &   200    \\\hline 
1QG Param \#        & 120          &   144   \\\hline 
2QG Param \#        & 40          &   48    \\\hline 


\end{tabular}
\vspace{0in}
\captionof{table}{Comparison of design overhead shows a 4-layer CopyQNN achieves $\sim$10$\times$ fewer queries with comparable hardware cost to QuantumLeak~\cite{Fu:IJCNN2024}.}
\label{tab:cmp_cost}
\end{minipage}
\vspace{-0.25in}
\end{figure*}  

\subsection{Quantum Contrastive Knowledge Transfer}
\textbf{Overview}.
The data-cleaning pipeline, while improving data quality, reduces the amount of available training data. Inspired by recent advancements in quantum transfer learning~\cite{wang2023hybrid} and quantum contrastive learning~\cite{jaderberg2022quantum}, \design~is designed to integrate both approaches, enabling efficient QNN extraction attacks even with limited training data.
As shown in Figure~\ref{f:overall_arch}(a), it comprises two primary quantum components: a quantum encoder (i.e., \texttt{QEnc}), which acts as a feature extractor, and a quantum classifier (i.e., \texttt{QClassifier}), which serves as the predictor. 
This approach leverages the principle that early layers in both classical~\cite{oquab2014learning} and quantum~\cite{wang2023hybrid} neural networks serve as general feature extractors, while the final layers are dedicated to task-specific predictions.
The training process starts with the \texttt{QEnc} being trained using contrastive learning on a local source dataset, as illustrated in Figure~\ref{f:overall_arch}(b). Once developed, the \texttt{QEnc} is transferred to the target domain and integrated with the \texttt{QClassifier} to form the overall model. 
The \texttt{QClassifier} is then trained using the limited cleaned dataset, as shown in Figure~\ref{f:overall_arch}(a).

\textbf{Data Augmentation}.
As conceptually show in Figure~\ref{f:overall_arch}(b), each input $I_i$ from the source dataset is transformed to generate two correlated data samples (i.e., $I_i{^1}$ and $I_i{^2}$) through a data augmentation module. 
The augmentation pipeline includes the following transformations: 
bilinear pooling to reduce the image size to 16x16, 
and random selection among \textit{Jitter} (adjusting contrast and brightness), \textit{Rotation} (rotating the image), \textit{Crop} (cropping a portion of the image), and \textit{Flip} (randomly flipping the image). 
Additionally, each transformation is tested with and without Gaussian blurring.

\textbf{Transferable Quantum Encoder}.
As shown in Figure~\ref{f:overall_arch}(b),
unlike classical contrastive learning~\cite{simclr, moco}, which require a \textit{projection head} after the encoder to map representations to a vector space, the learned quantum representations can be mapped to feature vectors (e.g., $Z_i^{1}$ and $Z_i^{2}$) via a quantum measurement layer, such as the Pauli-\texttt{Z} measurement.
Thanks to the advanced capability of operating on small batches with limited datasets, we adopt the Barlow Twins' objective function~\cite{barlow} as contrastive loss (i.e., $\mathcal{L}_{BT}$). 
For each minibatch of $N$ data, two augmented views for all data in the batch are obtained. These two batches of augmented data are then fed to \texttt{QEnc} to generate embeddings (e.g., $Z_i^1$ and $Z_i^2$).
Finally, $\mathcal{L}_{BT}$ is applied as follows:
\vspace{-4pt}\begin{equation}
\mathcal{L_{BT}} = \sum_{i} (1 - \mathbf{C}_{ii})^2 + \lambda \sum_{i} \sum_{j \neq i} \mathbf{C}_{ij}^2
\label{e:qnn_loss}
\vspace{-6pt}
\end{equation}
where $\mathbf{C}$ is the cross-correlation matrix, recording the similarity between the two outputs along the batch dimension. 
The first term in Equation~\ref{e:qnn_loss} encourages the diagonal elements of the cross-correlation matrix to approach 1, ensuring the embeddings remain invariant to augmentations. The second term drives the off-diagonal elements toward 0, decorrelating the different vector components of the embeddings.
The hyperparameter $\lambda$ (i.e., 0$\le$$\lambda$$\le$1) balances the importance of these two terms in the loss function.

\textbf{Adaptive Quantum Classifier}.
As illustrated in Figure~\ref{f:overall_arch}(a), we fix the pre-trained \texttt{QEnc} and integrate it with \texttt{QClassifier}, then train the task-specific quantum classifier using the cleaned queried dataset.

\subsection{NISQ Implementation}

\textbf{QNN Architecture}.
The \design~framework allows for flexible choices of QNN architecture without constraints. For simplicity, we adopt a commonly used QNN circuit ansatz, consisting of parameterized one-qubit gates (e.g., 1QG) for rotation followed by nearest-neighbor coupling of qubits using entanglement two-qubit gates (e.g., 2QG), as demonstrated in state-of-the-art QNNs\cite{Sukin2019_vqcc14}. This approach has shown superior expressive capability in various applications.
Figure~\ref{f:overall_arch}(c) shows the VQC architecture with a 4-qubit input as an example. Each VQC block features an \texttt{RZ} layer, an \texttt{RY} layer, an additional \texttt{RZ} layer, followed by a 2-qubit \texttt{CRX} entanglement layer. Both \texttt{QEnc} and \texttt{QClassifier} can use multiple VQC layers.


\textbf{Design Overhead}.
We focus on a binary classification using the MNIST dataset. QuantumLeak employs an ensemble of five QNNs, each consisting of two VQC layers, while~\design~eliminates the need for an ensemble, allowing for the construction of a deeper QNN with additional VQC layers under a similar total quantum resource budget. Specifically, \design~implements a \texttt{QEnc} with eight qubits to generate an 8-dimensional output feature vector and a \texttt{QClassifier} module with four qubits for binary classification. To maximize efficiency, \design~utilizes four VQC layers, as detailed in Table~\ref{tab:cmp_cost}, which compares the design overhead of~\design~and QuantumLeak~\cite{Fu:IJCNN2024}. Despite doubling the number of VQC layers, \design~incurs only a 20\% increase in hardware overhead while achieving a 90$\times$ reduction in queries. 

\textbf{Training Optimization}.
For the training of \texttt{QEnc}, each augmented input data is encoded into the 8-qubit QNN circuit using amplitude encoding. The training parameters are set to a batch size of 256, a learning rate of 5e-3, a weight decay of 1e-4, and 100 epochs. In training \texttt{QClassifier}, the Cross Entropy Loss is utilized, with a batch size that matches the number of queried images. Given the limited number of training samples, the learning rate is increased to 5e-2, and the training is conducted over 300 epochs.

\begin{figure*}[t!]
\begin{center}
\includegraphics[width=1\linewidth]{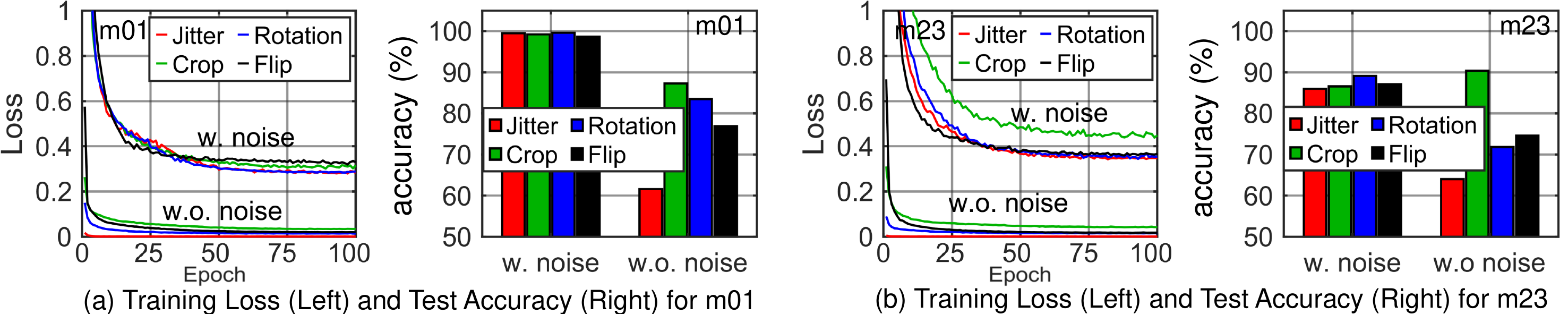}
\vspace{-0.15in}
\caption{Training loss and test accuracy with different data augmentation methods.}
\label{f:contras_performance}
\end{center}
\vspace{-0.3in}
\end{figure*}  

\vspace{0.15in}
\section{Experimental Methodology}
\label{sec:setup}
\vspace{-0.05in}

\textbf{Datasets}. 
We use MNIST and Fashion-MNIST datasets for evaluation.
For MNIST, we performed binary classification of 01 (\texttt{m01}), 23 (\texttt{m23}), 45 (\texttt{m45}), 67 (\texttt{m67}), and 89 (\texttt{m89}). 
For Fashion-MNIST, the tasks included \texttt{f01} (t-shirt/trouser), \texttt{f23} (pullover/dress), \texttt{f45} (coat/sandal), \texttt{f67} (shirt/sneaker), and \texttt{f89} (bag/ankle boot). 
Each task uses 3000 labeled images to train the cloud victim QNN and 1000 images for testing. For training \texttt{QEnc} with contrastive learning, we utilize an unlabeled subset from a different task, such as using \texttt{m01} data to pre-train \texttt{QEnc} for the \texttt{m23} task.

\textbf{NISQ Machines Configuration}.
We utilize BQSKit~\cite{Younis:BQST2021} for circuit synthesis and Qiskit~\cite{Qiskit} for deploying the synthesized circuits on NISQ computers.
All circuits were executed and measured on the 127-qubit \texttt{IBM\_Brisbane}. 
Queries to \texttt{IBM\_Brisbane} were evenly distributed across five different time points within 24 hours.

\textbf{Victim QNN}. 
For each classification task, we trained a victim QNN using four qubits. 
The victim QNN consists of one amplitude encoding layer, two repeated VQC layers, and one measurement layer. We used the QNN zoo from QuantumLeak~\cite{Fu:IJCNN2024} along with the QNN shown in Figure~\ref{f:overall_arch}(c) as potential victim QNN candidates.
We selected the negative log likelihood loss function for optimization and used the Adam optimizer with a learning rate of 5e-3 and a weight decay of 1e-4. The training sessions used a batch size of 32 and were conducted over 30 epochs.

\textbf{Evaluation Metrics}.
To evaluate the performance of the QNN model extraction attack, we use the inference accuracy of the learned substitute local QNN as a proxy metric. Higher inference accuracy of the substitute QNN indicates a more effective extraction attack.

\textbf{Schemes}. 
To compare~\design~with state-of-the-art QNN model extraction attacks, we established the following two baselines for each classification task:
\begin{itemize}[leftmargin=*, nosep, topsep=0pt, partopsep=0pt]
\item \textit{Base}: A local QNN using the same circuit architecture as the victim QNN is trained using all queried data points obtained across five time periods.

\item \textit{QLeak}: We adopted the QNN architecture with the best performance as reported in QuantumLeak~\cite{Fu:IJCNN2024}. The substitute QNN was trained using the ensemble learning method described in~\cite{Fu:IJCNN2024} with the same configuration.
\end{itemize}
We employed the Adam optimizer with a learning rate of 5e-3 and a weight decay of 1e-4. The batch size was set to 32. The training of local QNNs was conducted over 100 epochs.

\vspace{-0.2in}

\section{Evaluation and Results}
\label{sec:results}
\vspace{-0.05in}

We implement a practical~\design~with configuration in Table~\ref{tab:cmp_cost}. In the following, we first present results from the contrastive learning-based training to verify whether the proposed quantum contrastive knowledge transfer model can be successfully trained and to identify optimal data augmentation techniques and configuration parameters. 
We then analyze the impact of query rounds on the performance of the QNN extraction attack. 
Finally, we compare the overall performance across various \texttt{RR} values against the QuantumLeak attack. 
Note that, consistent with QuantumLeak, our results show that a local QNN using the same VQC ansatz as the victim QNN achieves the highest accuracy, and the following results are based on this configuration.

\textbf{Impact of Contrastive Learning Configuration}.
We conducted experiments to determine the optimal configuration for training the quantum encoder (\texttt{QEnc}). Since the VQC output is a feature vector rather than a predicted label, we focused on loss convergence as a key metric to evaluate the training process, while keeping the quantum classifier (\texttt{QClassifier}) fixed. 
As shown in Figure~\ref{f:contras_performance}, we use tasks \texttt{m01} and \texttt{m23} as examples to illustrate the training dynamics. The training loss exhibited gradual convergence across various data augmentation techniques, confirming the stability and effectiveness of the proposed approach. We considered scenarios both with and without the application of Gaussian noise. 
While the introduction of Gaussian noise led to a slight increase in loss, it did not hinder the overall training process. On the contrary, it enhanced the network’s ability to distinguish features within specific image classes, as evidenced by improved separability in the feature space, highlighted in the accompanying bar charts. Among the tested data augmentation methods—such as jitter, crop, rotation, and flip—no significant performance differences were observed. Based on these findings, we selected jitter in combination with Gaussian noise for the subsequent experiments, as this configuration effectively balances training stability and feature differentiation.

\begin{figure}[t!]
\centering
\vspace{-0in}
\includegraphics[width=.9\linewidth]{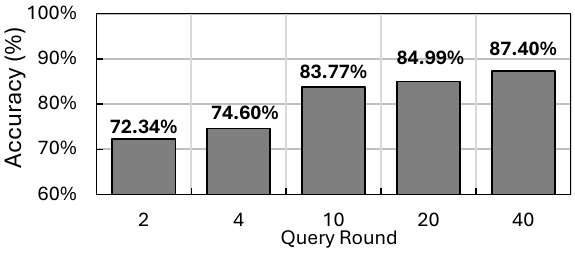}
\vspace{-0.05in}
\caption{Accuracy of~\design with different query rounds.}
\vspace{-0.3in}
\label{f:results_query_num}
\end{figure}

\textbf{Impact of Query Rounds}.
To evaluate the impact of query rounds on the effectiveness of the~\design~attack, we conducted experiments by systematically varying the number of query rounds for QNNaaS from 2 to 40, using the fixed data size reported in Table~\ref{tab:cmp_cost}. The resulting accuracy of the locally substituted QNN is presented in Figure~\ref{f:results_query_num}.
The results indicate that increasing the number of query rounds generally enhances the final accuracy by capturing more comprehensive noise features and refining the representation learned by the extracted model. For instance, increasing the query rounds from 4 to 10 results in a substantial accuracy improvement of 9.17\%, demonstrating that additional query rounds at this stage significantly enrich the extracted model's quality. However, as the number of query rounds grows further, the benefits diminish rapidly. For example, increasing the query rounds from 20 to 40 yields only a marginal accuracy gain of 2.41\%, indicating diminishing returns in extracting useful information beyond a certain threshold. 
It is important to highlight that higher query rounds come with trade-offs. While they improve the accuracy of the extracted model, they also elevate the risk of detection, thereby compromising the stealthiness of the model extraction attack. This trade-off underscores the importance of balancing query rounds to optimize performance while maintaining the attack’s covert nature.

\textbf{Overall Performance Comparison}.
We report the overall performance comparison of~\design~against two baseline schemes in Table~\ref{t:different_rem_rate}. The Remember Ratio (\texttt{RR}) is a critical parameter that determines the proportion of clean data included in the training set. 
To evaluate its impact, we gradually increased \texttt{RR} from 0.1, which uses only the top 10\% of high-variance labeled data, to 1, where all queried data are used for training without any data cleaning. The best and second-best performances for each task are highlighted in blue and teal, respectively.
The optimal \texttt{RR} setting was found to be 0.6, achieving an average performance improvement of 8.73\% over QuantumLeak across all tasks, with a particularly significant accuracy improvement of 29\% on task \texttt{m67}. In contrast, when \texttt{RR} is set to 1, where no data cleaning is applied, a substantial portion of the tasks exhibited worse performance compared to QuantumLeak, emphasizing the importance of effective data cleaning.
Overall, ~\design~demonstrates robust performance in training a local substitute QNN under varying NISQ conditions, achieving superior extraction attack accuracy compared to the state-of-the-art.


\begin{table*}[h]
\centering
\caption{Comprehensive performance comparison of
a na\"ive baseline (\textit{Base}), QuantumLeak (\textit{QLeak)}, and~\design~, where the best and second-best performances for each task are highlighted in {\color{blue}{blue}} and {\color{teal}{teal}}, respectively.}
\label{t:different_rem_rate}
\vspace{-0.05in}
\setlength{\tabcolsep}{7.4pt}

\begin{tabular}{|c|c|c|c|c|c|c|c|c|c|c|c|c|}
\hline
\multirow{2}{*}{\textbf{Task}}
& \multirow{2}{*}{\textbf{Base}}
& \multirow{2}{*}{\textbf{QLeak}~\cite{Fu:IJCNN2024}} 
& \multicolumn{10}{c|}{\textbf{\design~with Variable Remember Ratio (RR)}} \\\cline{4-13} 
& & & RR=0.1 & RR=0.2 & RR=0.3 & RR=0.4 & RR=0.5 & RR=0.6 & RR=0.7 & RR=0.8 & RR=0.9 & RR=1 \\ \hline\hline

\texttt{m01} & \color{blue}{99.6\%} & \color{blue}{99.6\%} & 91.2\% & 81.4\% & 72.8\% & 92.5\% & 97.6\% & \color{teal}{98.7\%} & 98.0\% & 98.4\% & 98.4\% & 98.4\% \\ \hline
\texttt{m23} & 50.8\% & 50.0\% & \color{blue}{83.9\%} & \color{teal}{81.3\%} & 76.8\% & 63.3\% & 62.2\% & 67.9\% & 64.9\% & 59.8\% & 63.5\% & 70.0\% \\ \hline
\texttt{m45} & 60.0\% & 64.7\% & 42.7\% & 47.4\% & 50.1\% & 65.4\% & 48.6\% & \color{blue}{71.8\%} & \color{teal}{69.9\%} & 67.4\% & 62.2\% & 64.1\% \\ \hline
\texttt{m67} & 87.9\% & 59.9\% & 75.9\% & 79.5\% & 73.8\% & 78.4\% & 84.3\% & \color{blue}{88.9\%} & 87.5\% & 88.3\% & \color{teal}{88.4\%} & 83.9\% \\ \hline
\texttt{m89} & 50.0\% & 50.0\% & 43.3\% & 58.7\% & 52.9\% & 53.1\% & 59.7\% & \color{blue}{68.1\%} & 58.4\% & 64.2\% & \color{teal}{64.8\%} & 62.1\% \\ \hline
\texttt{f01} & 63.0\% & 52.4\% & 46.1\% & 49.5\% & \color{teal}{82.3\%} & 77.5\% & \color{blue}{88.7\%} & 79.2\% & 74.4\% & 74.1\% & 68.4\% & 69.9\% \\ \hline
\texttt{f23} & \color{blue}{90.0\%} & 84.5\% & 60.6\% & 55.9\% & 51.9\% & 59.9\% & 71.4\% & 63.9\% & 80.0\% & 79.5\% & 83.5\% & \color{teal}{89.8\%} \\ \hline
\texttt{f45} & 93.6\% & 93.6\% & 50.0\% & 50.0\% & 81.3\% & 81.4\% & 93.9\% & 94.8\% & 94.5\% & \color{blue}{95.1\%} & \color{teal}{95.0\%} & 94.1\% \\ \hline
\texttt{f67} & \color{teal}{96.1\%} & \color{blue}{97.9\%} & 50.9\% & 49.3\% & 91.1\% & 91.7\% & 88.5\% & 89.1\% & 90.1\% & 94.0\% & 92.8\% & 92.9\% \\ \hline
\texttt{f89} & 50.0\% & 50.0\% & 60.2\% & 63.0\% & 65.9\% & \color{blue}{70.9\%} & \color{teal}{70.6\%} & 67.5\% & 63.1\% & 63.1\% & 65.9\% & 69.9\% \\ \hline
\end{tabular}
\vspace{-0.2in}
\end{table*}

\section{Conclusion}
This work investigates QNN model extraction attacks on NISQ computers under varying noise conditions. We propose the~\design~framework, which refines queried datasets via data preprocessing to remove mislabeled responses from QNN-as-a-Service servers, then applies quantum-domain contrastive and transfer learning to efficiently train a substitute QNN. Experimental results show that\design~achieves superior performance with significantly fewer queries, outperforming state-of-the-art extraction methods.

\vspace{-2pt}
\section*{Acknowledgment}
This work was supported in part by NSF OAC-2417589 and NSF CNS-2143120. 
We thank the IBM Quantum Researcher \& Educators Program for their support of Quantum Credits.

\vspace{-2pt}
\bibliographystyle{ieeetr}
\bibliography{reference,quantum}

\end{document}